\documentclass[twocolumn]{aastex631}
\pdfoutput=1

\usepackage[separate-uncertainty=true, range-phrase=\text{-}, range-units=single,
angle-symbol-over-decimal, retain-explicit-plus, multi-part-units=single]{siunitx}
\usepackage{amsmath}
\usepackage{graphicx}
\usepackage{chemformula}
\usepackage{acro}
\usepackage{layouts}
\usepackage{float}
\usepackage{multirow}
\acsetup{patch/longtable=false}
\usepackage{lipsum}
\usepackage[T1]{fontenc}

\received{}

\shorttitle{}
\shortauthors{}

\DeclareSIUnit{\mag}{mag}
\DeclareSIUnit{\pixel}{pixel}
\DeclareSIUnit{\parsec}{pc}
\DeclareSIUnit{\arcsec}{arcsec}
\DeclareSIUnit{\arcmin}{arcmin}
\DeclareSIUnit{\solarlum}{\mbox{$\mathcal{L}_\odot$}}
\DeclareSIUnit{\solarmass}{\mbox{$\mathcal{M}_\odot$}}
\DeclareSIUnit{\solarmetal}{\mbox{$Z_\odot$}}
\DeclareSIUnit{\year}{yr}
\DeclareSIUnit{\deg}{deg}
\DeclareSIUnit{\erg}{erg}
\DeclareSIUnit{\dex}{dex}
\DeclareSIUnit{\angstrom}{\textup{\AA}}
\DeclareSIUnit{\radius}{\mbox{$R_{25}$}}
\DeclareSIUnit{\spaxel}{spaxel}

\newcommand{\ha}{H$\alpha$}
\newcommand{\hb}{H$\beta$}
\newcommand{\oiii}{[\ion{O}{3}]}
\newcommand{\nii}{[\ion{N}{2}]}
\newcommand{\oii}{[\ion{O}{2}]}
\newcommand{\sii}{[\ion{S}{2}]}

\newcommand{\hii}{\ion{H}{2}}
\newcommand{\hi}{\ion{H}{1}}

\newcommand{\ppxf}{\textsc{ppxf}}

\DeclareAcroEnding{possessive}{'s}{'s}

\NewAcroCommand\acg{m}{\acropossessive\UseAcroTemplate{first}{#1}}

\DeclareAcronym{sfr}{short=SFR, long=star formation rate}
\DeclareAcronym{ssfr}{short=sSFR, long=specific star formation rate}
\DeclareAcronym{psf}{short=PSF, long=point spread function}
\DeclareAcronym{agn}{short=AGN, long=active galactic nucleus, long-plural-form=active galactic nuclei}
\DeclareAcronym{lsb}{short=LSB, long=low surface brightness}
\DeclareAcronym{ism}{short=ISM, long=interstellar medium}
\DeclareAcronym{sdss}{short=SDSS, long=Sloan Digital Sky Survey}
\DeclareAcronym{dig}{short=DIG, long=diffuse ionized gas}
\DeclareAcronym{ifs}{short=IFS, long=integral field spectroscopy}
\DeclareAcronym{imf}{short=IMF, long=initial mass function}
\DeclareAcronym{sfh}{short=SFH, long=star formation history, long-plural-form=star formation histories}
\DeclareAcronym{sed}{short=SED, long=spectral energy distribution}
\DeclareAcronym{pings}{short=PINGS, long=PPak IFS Nearby Galaxies Survey}
\DeclareAcronym{sn}{short=S/N, long=signal-to-noise}
\DeclareAcronym{pmas}{short=PMAS, long=Potsdam Multi-Aperture Spectrophotometer}
\DeclareAcronym{cavex}{short=CAVEX, long=Calar Alto Extinction monitor}
\DeclareAcronym{twomass}{short=2MASS, long=Two Micron All Sky Survey}
\DeclareAcronym{bulge}{short=PB, long=putative bulge}

\newcolumntype{P}[1]{>{\centering\arraybackslash}p{#1}}

\begin{document}

\title{When Is a Bulge Not a Bulge? Revealing the Satellite Nature of NGC~5474's Bulge}
\author{Ray Garner, III}
\affiliation{Department of Physics and Astronomy, Texas A\&M University, 578 University Dr., College Station, TX, 77843, USA}
\affiliation{George P.\ and Cynthia W.\ Mitchell Institute for Fundamental Physics \& Astronomy, Texas A\&M University, 578 University Dr., College Station, TX, 77843, USA}

\author{J.\ Christopher Mihos}
\affiliation{Department of Astronomy, Case Western Reserve University, 10900 Euclid Ave., Cleveland, OH 44106, USA}

\author{F.\ Fabi\'{a}n Rosales-Ortega}
\affiliation{Instituto Nacional de Astrof\'{\i}sica, \'{O}ptica y Electr\'{o}nica, Luis Enrique Error 1, 72840 Tonantzintla, Mexico}

\correspondingauthor{Ray Garner, III}
\email{ray.three.garner@gmail.com}

\begin{abstract}
A satellite galaxy of the nearby spiral M101, NGC~5474 has a prominent bulge offset from the kinematic center of the underlying star-forming disk that has gained attention in recent years. Recent studies have proposed that this putative offset bulge is not a classical bulge within the plane of the disk but instead a dwarf companion galaxy along the line-of-sight. Using \acl{ifs} data taken as part of the \acf{pings}, we perform the first analysis of the stellar and gas kinematics of this \acl{bulge} and portions of the disk. We find a radial velocity offset of $\sim$\SI{24}{\kilo\metre\per\second} between the emission lines produced by the disk \hii\ regions and the absorption lines produced by the \acl{bulge} stellar component. We interpret this velocity offset as evidence that the \acl{bulge} and disk are two separate objects, the former orbiting around the latter, supporting simulations and observations of this peculiar system. We attempt to place this external companion into the context of the M101 Group and the M101-NGC~5474 interaction. 
\end{abstract}

\section{Introduction}

Located within the M101 Group at a distance of \SI{6.9}{\mega\parsec} (see \citealt{matheson2012} and references therein) lies NGC~5474, a peculiar star-forming galaxy. In this sparsely populated group, NGC~5474 is the largest and brightest ($R_{25} = \ang{;2.4;}$, $M_B \simeq -17.9$; \citealt{devaucouleurs1991}) satellite galaxy of M101 (NGC~5457), which is itself a peculiar galaxy. NGC~5474 is also relatively close to M101 with a projected angular separation of \ang{;44;} corresponding to a physical separation of \SI{88}{\kilo\parsec}. This relatively close separation and the lack of other massive galaxies in the M101 Group has led many authors to propose that a recent interaction between M101 and NGC~5474 is responsible for producing the peculiarities seen in M101 \citep[e.g.,][]{beale1969,waller1997,mihos2012,mihos2013,mihos2018,xu2021,linden2022,garner2022,garner2024}. 

However, just like M101, NGC~5474 is home to numerous asymmetries and oddities remarked upon over the years \citep{kornreich1998}. Early observations focused on its \hi\ disk and found a smoothly varying disk in the central region out to \SI{5}{\kilo\parsec}, beyond which lie distortions with a change in position angle by as much as $\simeq$\ang{50} \citep{rownd1994}. The \hi\ disk of NGC~5474 is also connected to the southwestern edge of M101 via a bridge of intermediate velocity \hi\ gas \citep{huchtmeier1979,vanderhulst1979}, interpreted as tidal debris from the M101-NGC~5474 interaction \citep{mihos2012}. However, researchers have not found any optical debris down to extremely low surface brightnesses, $\mu_V \sim 28$ \citep{mihos2013,garner2021}.

\begin{figure*}
\includegraphics[keepaspectratio,width=0.5\textwidth]{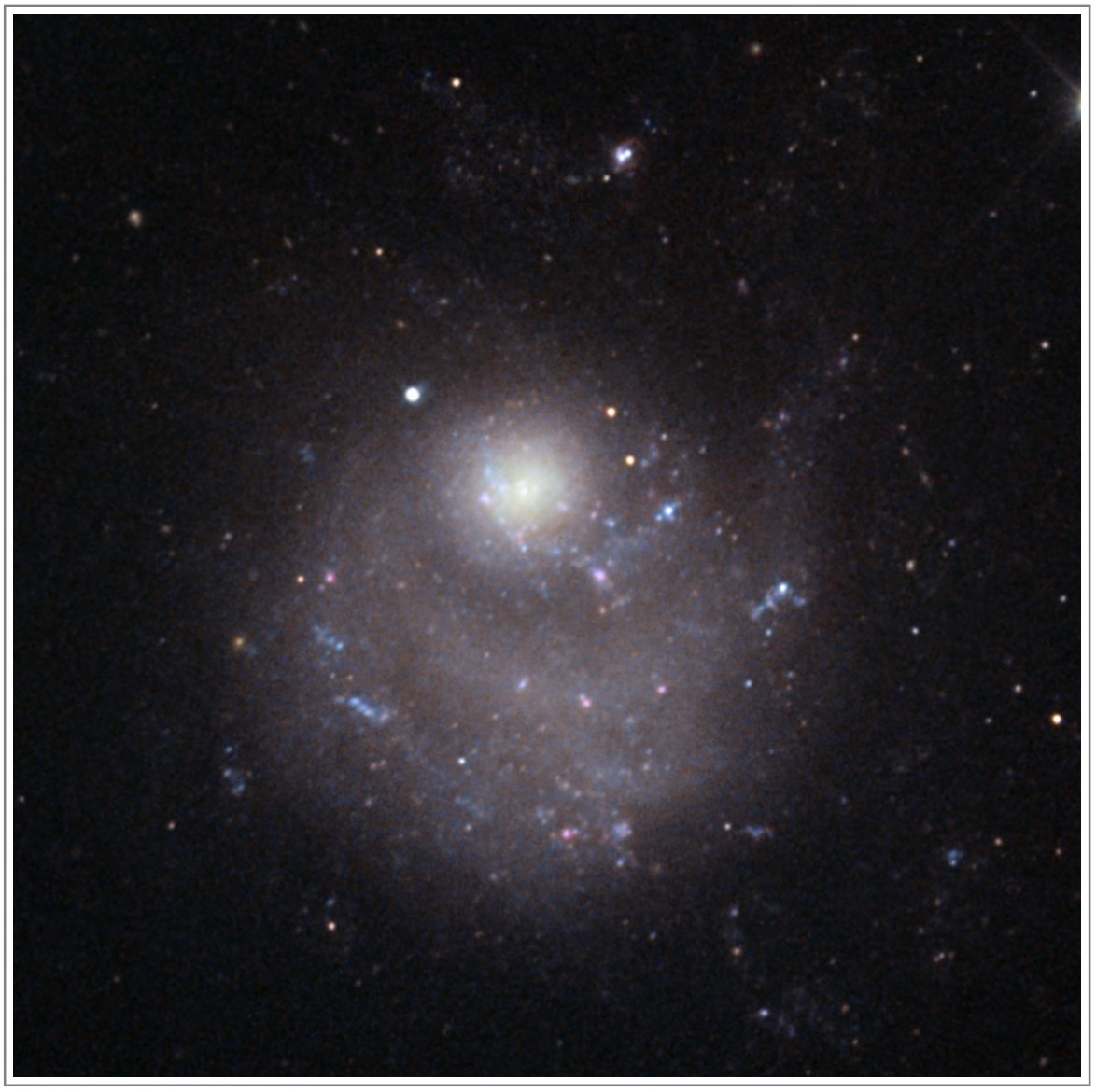}
\includegraphics[keepaspectratio,width=0.5\textwidth]{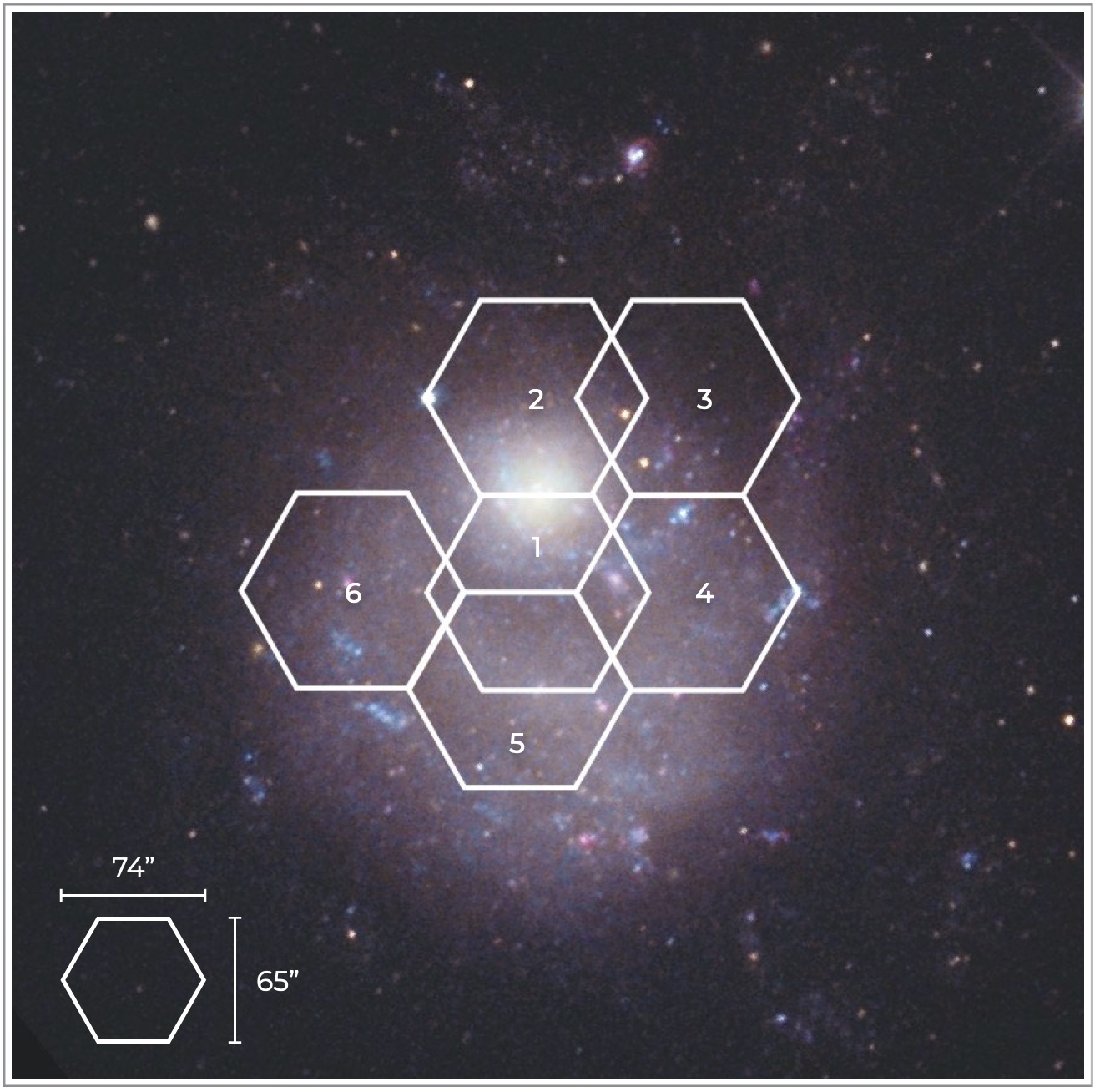}
\caption{An RGB color image of NGC~5474 with (right) and without (left) overlaid PPak pointings. Both images measure $\ang{;6;} \times \ang{;6;}$. North is up and east is to the left. Credits: KPNO/NOIRLab/NSF/AURA/Adam Block.}
\label{rgb}
\end{figure*}

Optically, NGC~5474 appears as a typical dwarf spiral galaxy with loose spiral arms, circular outer isophotes \citep{mihos2013}, and numerous sites of star formation, including an extended UV disk \citep{thilker2007}. The left panel of Figure~\ref{rgb} illustrates the exception: an optical bulge at the northern edge of a nearly face-on stellar disk. The bulge does not appear strikingly offset with respect to the overall spiral pattern of the disk but instead shows a \SI{1}{\kilo\parsec} offset north from the kinematic center of NGC~5474's \hi\ and \ha\ disks \citep{rownd1994,kornreich2000,epinat2008}. 

Historically, the offset bulge has been gathered together with the \hi\ distortions and bridge as being caused by the interaction NGC~5474 had with M101 $\sim$\SI{300}{\mega\year} ago \citep{mihos2013,mihos2018,linden2022}. However, recent work has called the nature of this bulge into question. \cite{bellazzini2020} renewed interest in NGC~5474, finding that the \acl{bulge} (\acs{bulge} hereafter, adopting the nomenclature of \citealt{pascale2021}) has many structural properties similar to dwarf galaxies, especially M110 (NGC~205), a dwarf elliptical satellite galaxy of M31. They proposed that NGC~5474 also bears the signs of a recent interaction with a dwarf companion and pointed to the \ac{bulge} as the culprit. \cite{pascale2021} used $N$-body hydrodynamical simulations to investigate this possibility and found that it is highly unlikely that the \ac{bulge} resides in the plane of the galaxy. Suppose, instead, the \ac{bulge} was an early-type satellite galaxy on a polar orbit around NGC~5474. In that case, simple projection effects can produce the apparent offset, and the interaction between the two bodies explains the warped \hi\ disk.  Most recently, \cite{bortolini2024} investigated the \acp{sfh} of the \ac{bulge}, disk, and an over-density of stars to the southwest and found a synchronized burst of activity around \SIrange{10}{35}{\mega\year} ago, which they interpret as a signature of a more recent interaction between the disk and \ac{bulge}. 

Despite the growing body of evidence that suggests the disk of NGC~5474 and the \ac{bulge} are two separate objects, no conclusive evidence in the form of a radial line-of-sight velocity difference or offset has been shown. Unfortunately, all of the available stellar velocity fields are based on emission lines that trace the star-forming disk \citep{ho1995,epinat2008,moustakas2010}, while the \ac{bulge} is dominated by old and intermediate-age stars with kinematics best traced by absorption line spectra. Thus, we need spatially resolved \ac{ifs} data sets that provide both emission line and absorption line measurements. This type of data has been used in the past to distinguish between relaxed virialized systems and merger events \citep[e.g.,][]{flores2006,shapiro2008,bellocchi2012,bellocchi2013,torres-flores2014,oh2022}, including merger stages \citep{barreraballesteros2015}, and measuring a radial velocity offset would be relatively simple to perform. 

\cite{bellazzini2020} proposed and attempted exactly this methodology. Due to the limitations of their data, they could only conclude that the radial velocity difference between the emission and absorption within the \ac{bulge} is $\lesssim$\SI{50}{\kilo\metre\per\second}. This small velocity offset rules out a chance superposition of kinematically unrelated systems, but the critical question of the nature of the \ac{bulge} was left unanswered. In this paper, we attempt to conclusively measure a velocity offset between the disk and \ac{bulge} of NGC~5474 using \ac{ifs} data acquired as part of the \ac{pings}. Owing to the generous wavelength range and large field-of-view, we can extract two-dimensional velocity maps for both the stellar and ionized gas components across much of the galaxy. These maps' velocity differences should answer the outstanding question of NGC~5474's peculiar \acl{bulge}.

\section{Observations and Data Reduction}

The spectroscopic observations of NGC~5474 are part of the \acl{pings} (\acs{pings}; \citealt{rosalesortega2010}), a project aimed at constructing 2D spectroscopic mosaics for a sample of nearby spiral galaxies. The \ac{pings} observations were conducted using the \SI{3.5}{\metre} telescope at the Calar Alto Observatory with the \acl{pmas} (\acs{pmas}; \citealt{roth2005}) in PPak mode \citep{verheijen2004,kelz2006}. This mode employs a retrofitted bare bundle of 331 optical fibers, sampling the target with a spatial resolution of \ang{;;2.7} per fiber over a hexagonal area with a $\ang{;;74} \times \ang{;;65}$ footprint and a \SI{65}{\percent} filling factor. The sky background is sampled by 36 additional fibers arranged in 6 mini-IFU bundles of 6 fibers each, distributed in a circular pattern at $\sim$\ang{;;90} from the center and at the edges of the central hexagon. Additionally, 15 fibers are illuminated by internal lamps for calibration purposes. The instrument was configured with the V300 grating, covering a wavelength range of \SIrange{3700}{7100}{\angstrom} with a spectral resolution of $\sim$\SI{10}{\angstrom} FWHM, corresponding to a velocity $\sim$\SI{460}{\kilo\metre\per\second} for \ha. While this FWHM velocity resolution is rather high, our subsequent analysis using \ppxf\ (Section~\ref{sec:analysis}) provides much higher velocity accuracy on the fitted spectra (see also \citealt{cappellari2017}, Section~4).

Different observing strategies were employed depending on the size of the \ac{pings} galaxies, which ranged from $\sim$\ang{;1;} to \ang{;10;} in diameter. By construction, the initial exposure was positioned at a predefined geometrical position, which, contingent on the galaxy's morphology or the selected mosaicing pattern, may or may not coincide with the galaxy's bright bulge. Subsequent pointings generally followed a hexagonal pattern, aligning the mosaic pointings with the shape of the PPak science bundle. Each pointing center was radially offset by \ang{;;60} from the previous one. Due to the shape of the PPak bundle and the design of the mosaics, 11 spectra at the edge of each hexagon overlapped with the same number of spectra from the preceding pointing. This strategy was chosen to maximize the covered area while ensuring sufficient overlap to align the exposures taken under varying atmospheric conditions and/or at different epochs. 

In the case of NGC~5474, the original strategy was to observe the galaxy with a standard mosaic configuration comprising one central position and one concentric ring. This configuration would cover the optical area of the galaxy ($\sim$$\ang{;4.8;} \times \ang{;4.3;}$). However, due to the distorted morphology of NGC~5474, the position of the central pointing was selected to ensure that the entire mosaic would encompass the optical area of the galaxy symmetrically. This configuration implied a \ang{;;30} offset in declination (toward the south) of the central position with respect to the bright pseudo-bulge of the galaxy. The initial two positions were observed in June~2008, employing this scheme. However, positions 3 to 6 were observed in service mode in August~2008. The previous central coordinates of position 1 were not properly recovered, and the bright pseudo-bulge was chosen as the reference for the mosaicing during this run, resulting in the odd mosaic scheme depicted in the right panel of Figure~\ref{rgb}. The alignment of the pointings was refined during the final data reduction using broadband images (see below).

All positions were observed in dithering mode, taking three dithered exposures per position. Positions 1 and 2 were repeated during a third run in April~2009 due to quality issues in the initial observations. For position 6, the dithering was incomplete due to the low altitude of the object. The acquisition time per PPak field in dithering mode was $2 \times \SI{600}{\second}$ per dithering frame (i.e., \SI{60}{\minute} of exposure per position). The average seeing was \ang{;;1.3} with a maximum of \ang{;;1.5} over the three observing runs (below the fiber size). The total exposure time for the \ac{pings} observations of NGC~5474 amounts to \SI{6}{\hour}. These exposure times provided spectroscopy with $\text{S/N} \geq 20$ in the continuum and $\text{S/N} \geq 50$ in the \ha\ emission line for the brightest \hii\ regions. In total, \num{5958} individual spectra were obtained for this galaxy. 


The reduction of the \ac{pings} observations for NGC~5474 followed the standard steps for fiber-based \ac{ifs}. Pre-reduction processing was performed using standard \textsc{iraf}\footnote{IRAF is distributed by the National Optical Astronomy Observatories, which are operated by the Association of Universities for Research in Astronomy, Inc., under cooperative agreement with the National Science Foundation.} packages, while the main reduction was performed using the \textsc{r3d} software for fiber-fed and \ac{ifs} data \citep{sanchez2006} in combination with the \textsc{e3d} and \textsc{pingsoft} \ac{ifs} visualization and manipulation software \citep{sanchez2004,rosalesortega2011}. This process produces distortion- and transmission-corrected, sky-subtracted, wavelength- and relative flux-calibrated spectra. An additional spectrophotometric calibration correction was applied by comparing the \ac{ifs} data with $B$, $V$, $R$, and \ha\ imaging photometry from the SINGS legacy survey \citep{kennicutt2003}. The estimated spectrophotometric accuracy of the \ac{ifs} mosaic is approximately \SI{0.2}{\mag}. During this re-normalization process, the relative astrometric accuracy between the pointings of the \ac{ifs} mosaic was improved to a \mbox{$\sim$\ang{;;0.3}} level based on the rms of the centroid differences of foreground stars. Finally, the fiber-based \ac{ifs} data for NGC~5474 were spatially resampled into a data cube with a regular grid of \ang{;;1} \si{\per\spaxel} using a flux-conserving, natural neighbor, non-linear interpolation method as described in \cite{sanchez2012} developed for the CALIFA survey. A detailed explanation of the observing strategy and data reduction can be found in \cite{rosalesortega2010}.

During the analysis detailed in the next section, we noticed a systematic offset in the sky coordinates of the final data cube. Using the centroids of two bright stars from the \ac{twomass} All-Sky Catalog of Point Sources \citep{skrutskie2006} that are also in the PPak FOV, we measured the offset and found that the PPak coordinates are shifted \ang{;;5} to the northwest. In the remainder of this paper, we have either corrected any reported coordinates for this offset or used relative coordinates.

\section{Data Analysis}\label{sec:analysis}

In order to determine the nature of the \acl{bulge} of NGC~5474, we relied on the distinct stellar populations that make up the disk and \ac{bulge} and their resulting spectra. Old and intermediate-age stars primarily produce the light of the offset \ac{bulge}, while young stellar populations and gaseous \hii\ regions contribute to the disk light \citep{bellazzini2020,bortolini2024}. Thus, where the \ac{bulge} dominates the light profile, the absorption lines present in the stellar continuum should trace its kinematics. In contrast, emission lines produced by the \hii\ regions trace the disk kinematics. If the \ac{bulge} is kinematically separate from the disk, the two spectral components should be offset according to their radial velocities. This is the technique proposed (and attempted) by \cite{bellazzini2020}. 

We used two methods to measure the radial velocity offset between the disk and \ac{bulge}. The first method sums the total light within a series of circular apertures centered on the \ac{bulge} to measure the radial velocities of the absorption and emission lines. This technique has the benefit of increasing the \acs{sn} of the data. The second method utilized our entire FOV to produce two-dimensional spatially-resolved velocity maps. This method necessarily reduces the \ac{sn} but puts the \ac{bulge} into dynamical context with the rest of the disk. However, both techniques give the same answer: a radial velocity offset of $\simeq$\SI{24}{\kilo\metre\per\second} between the disk and \ac{bulge}. We detail both methods in the following subsections. 

Since both methods rely on the Python version of the penalized pixel-fitting code \ppxf\ (v9.1.1; \citealt{cappellari2004,cappellari2017,cappellari2023}) to fit the spectra and extract radial velocities, we detail our setup here. The \ppxf\ code requires the user to select a base set of stellar population templates to construct the final model. We adopted the \cite{bruzual2003} high-resolution model templates for this analysis. These models adopt a \cite{salpeter1955} IMF and the Padova isochrones \citep{marigo2008}. We broaden these templates to match the spectral resolution of the \ac{pings} data using the \texttt{log\_rebin} code provided with the \ppxf\ package. During the fitting process, we use tenth-order additive polynomials and no multiplicative polynomials. In the following subsections, we remark on how we apply \ppxf\ specifically in each method.


\subsection{Circular Apertures on the Putative Bulge}\label{sec:circ-aps}

\begin{figure}
\includegraphics[keepaspectratio,width=\columnwidth]{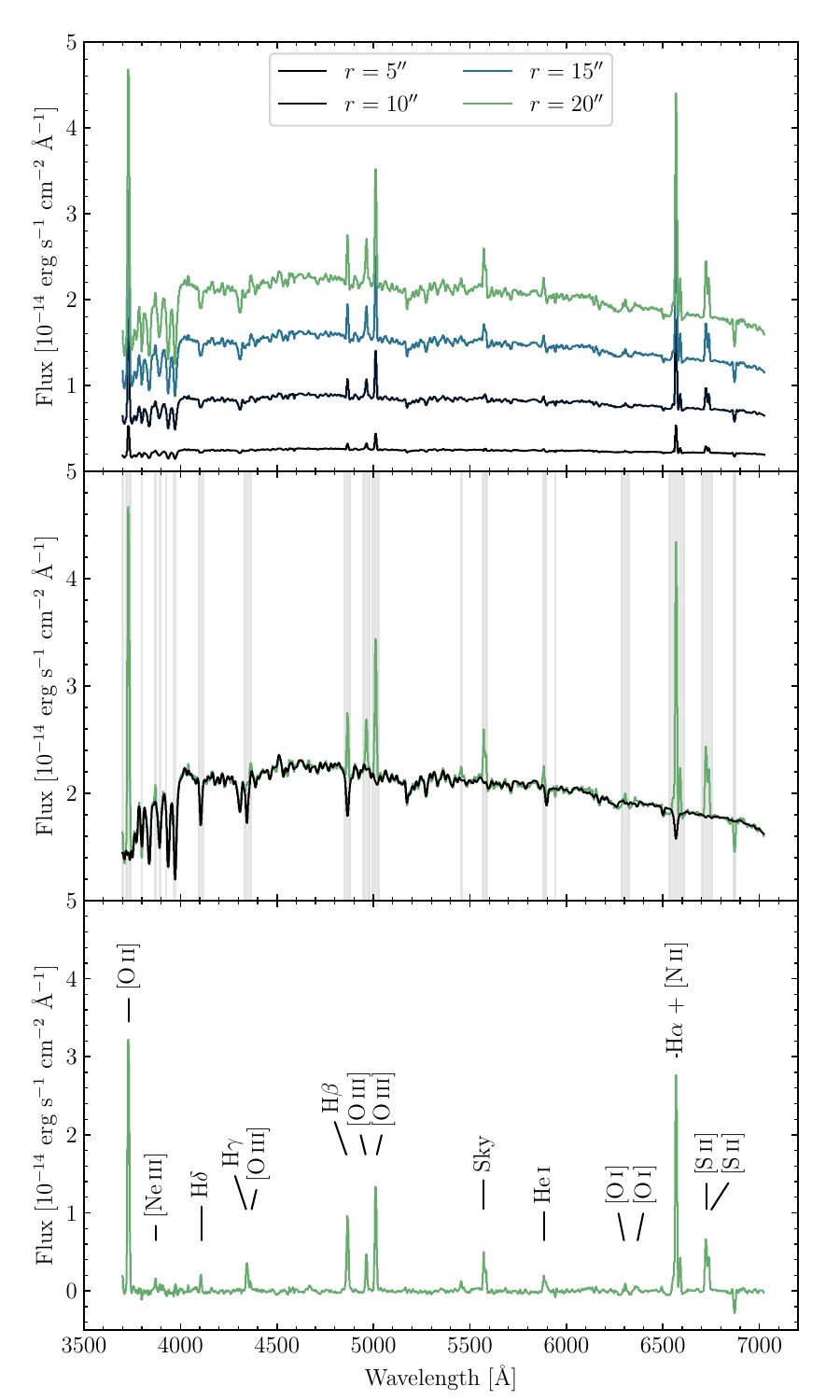}
\caption{The spectra in circular apertures centered on the bulge of NGC~5474. The top panel shows the spectra for the four apertures used with their radii indicated in the legend. The middle panel shows an example of the \ppxf\ fit to the stellar continuum overlayed in black for the \ang{;;20} spectrum. The light gray bands correspond to the spectral regions masked during the stellar continuum fitting. The bottom panel shows the residual after subtracting the model from the original spectrum; the detected emission lines are labeled.}
\label{aperture_spectra}
\end{figure}

This method extracts spectra summed over circular apertures centered at the location of the \ac{bulge}. The radius of the aperture is a trade-off between disk contamination and total \ac{sn}. Smaller apertures minimize contamination from the disk stellar light but yield lower \ac{sn}, while bigger apertures maximize the total light and \ac{sn} but add more disk stellar light. Therefore, we extracted the total spectrum in four apertures with radii varying from \ang{;;5} to \ang{;;20} in \ang{;;5} increments; beyond \ang{;;20} the data quality decreases. The top panel of Figure~\ref{aperture_spectra} shows the resulting spectra. Each aperture contains absorption lines from the \ac{bulge} and emission lines from \hii\ regions in the disk. 

To understand the origin of the changing properties with aperture size, we need to quantify the fraction of light contributed by the \ac{bulge} within our apertures. We adapted the method from \cite{bellazzini2020}, using the $V$-band image from \cite{mihos2013} to derive the radial flux profile of the \ac{bulge} and the surrounding disk. We used a series of circular apertures extending out to \ang{;;145}, where within each aperture we determined the median flux value and fitted the resulting profile. To estimate the contribution of the underlying disk, we calculated the average flux from the radial light profile. We calculated the radial fraction of light attributed to the \ac{bulge} using these values. Table~\ref{bootstrap_table} lists the fraction of the \ac{bulge} light, $f_{\text{PB}}$, in each circular aperture. As expected, the \ac{bulge} dominates the innermost \ang{;;5} aperture with $f_{\text{PB}} = \SI{88.1}{\percent}$, while the fraction decreases to $f_{\text{PB}} = \SI{77.0}{\percent}$ in the \ang{;;20} aperture. 


We apply \ppxf\ twice to each spectrum. The first pass is to fit only the stellar velocity and velocity dispersion. During this pass, we mask the emission lines and outliers using the method outlined in Section~6.5 of \cite{cappellari2023}. In order to estimate the uncertainties on the resulting parameters, we used the wild bootstrapping method of \cite{davidson2008} and apply it 500 times. In short, this is the same as standard residual bootstrapping, but the residual for each data point is randomly multiplied by $+1$ or $-1$ with a probability of $0.5$ before resampling in order to account for heteroskedasticity in the data. The middle panel of Figure~\ref{aperture_spectra} shows an example fit to the stellar continuum of the \ang{;;20} aperture spectrum. We automatically mask the regions in the light gray bands during the fitting process. 

After fitting for the stellar continuum and its kinematics, the second pass of \ppxf\ fixes constant the stellar continuum model and fits for the emission lines. We take the median stellar kinematics as the fixed values. Again, we apply the wild bootstrapping method $500$ times to estimate the uncertainties in the gas kinematics. The bottom panel of Figure~\ref{aperture_spectra} shows an example of the emission line spectrum in the \ang{;;20} aperture after subtracting the best stellar continuum model from the data. Various emission lines are labeled. 

\begin{figure*}[htb!]
\includegraphics[keepaspectratio,width=\textwidth]{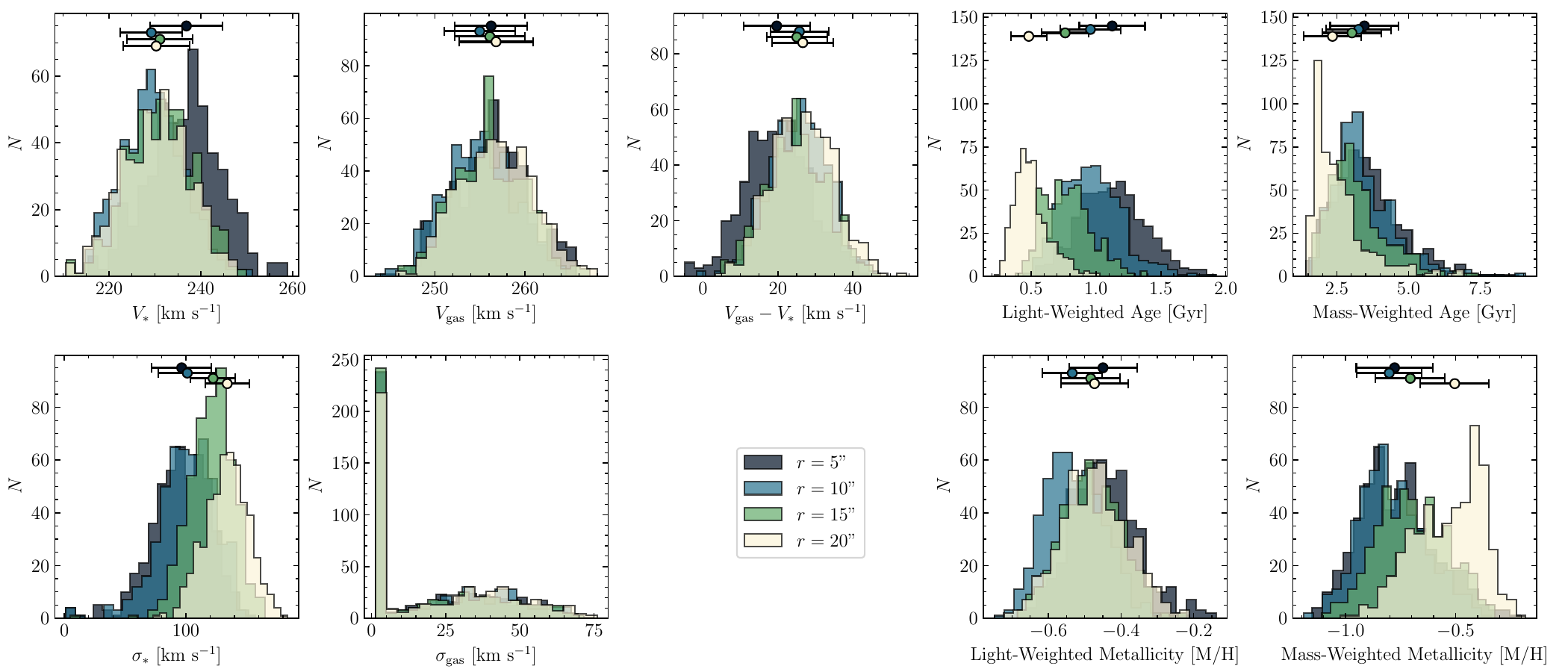}
\caption{The histograms of our bootstrapping procedure for each aperture size. From top to bottom, left to right: stellar velocity, stellar velocity dispersion, gas velocity, gas velocity dispersion, $V_{\text{gas}} - V_\ast$ velocity difference, light-weighted age and metallicity, and mass-weighted age and metallicity. In each panel, the different aperture sizes are indicated by different colors, see the legend in the bottom-center. The points with error bars at the top of each panel indicates the median and standard deviation of the histograms. Those values are reported in Table~\ref{bootstrap_table}. For the gas velocity dispersion, the dispersions often failed to converge, leading to the unphysical peak at small dispersion values. For this reason, we do not calculate the mean and standard deviation for the gas velocity dispersion.}
\label{bootstrap_hist}
\end{figure*}

Histograms showing the bootstrapped values of the stellar and gas kinematics, and both light-weighted and mass-weighted ages and metallicities are shown in Figure~\ref{bootstrap_hist}. The median values are reported in Table~\ref{bootstrap_table}. In addition to the median values, we quantify the spread of these histograms with the standard deviation, reported in Table~\ref{bootstrap_table} as the uncertainties on each measurement. We stress that these are not the uncertainties on any individual measurement of a particular property. Previous work using simulated spectra at a resolution comparable to the \ac{pings} data has explored the true uncertainties on a single fit \citep{marmolqueralto2011,sanchez2016}. For instance, \cite{marmolqueralto2011} found a velocity recovery error of $\lesssim$\SI{5}{\percent}. Thus, an error of $\simeq$\SI{10}{\kilo\metre\per\second} on any individual fit of the stellar velocity is likely present. These individual measurement uncertainties do not impact our broad conclusions below since relative velocity differences are typically better constrained than absolute velocity measurements as the overall instrumental systematic uncertainties tend to cancel out. We refer the interested reader to the aforementioned papers for more information.

\begin{deluxetable*}{l c c c c}[hbt!]
\tablecaption{Median Properties of the Bootstrapped Apertures \label{bootstrap_table}}
\tablehead{\colhead{Property} & \colhead{\ang{;;5}} & \colhead{\ang{;;10}} & \colhead{\ang{;;15}} & \colhead{\ang{;;20}}} 
\startdata
Bulge Light Fraction, $f_{\text{PB}}$ & \SI{88.1}{\percent} & \SI{85.3}{\percent} & \SI{81.6}{\percent} & \SI{77.0}{\percent} \\
Stellar Velocity, $V_\ast$ [\si{\kilo\metre\per\second}] & \num{236.9 \pm 7.9} & \num{229.2 \pm 6.7} & \num{231.1 \pm 7.1} & \num{230.3 \pm 7.3} \\
Stellar Velocity Dispersion, $\sigma_\ast$ [\si{\kilo\metre\per\second}] & \num{96.6 \pm 24.6} & \num{101.2 \pm 23.6} & \num{122.4 \pm 18.4} & \num{134.1 \pm 18.0} \\
Gas Velocity, $V_\text{gas}$ [\si{\kilo\metre\per\second}] & \num{256.2 \pm 4.0} & \num{255.0 \pm 3.9} & \num{256.1 \pm 3.9} & \num{256.8 \pm 4.1} \\
Light-Weighted Age [\si{\giga\year}] & \num{1.1 \pm 0.3} & \num{1.0 \pm 0.2} & \num{0.8 \pm 0.2} & \num{0.5 \pm 0.1} \\
Mass-Weighted Age [\si{\giga\year}] & \num{3.5 \pm 1.2} & \num{3.3 \pm 1.1} & \num{3.0 \pm 1.0} & \num{2.3 \pm 1.0} \\
Light-Weighted Metallicity [M/H] & \num{-0.45 \pm 0.09} & \num{-0.53 \pm 0.08} & \num{-0.48 \pm 0.08} & \num{-0.47 \pm 0.09} \\
Mass-Weighted Metallicity [M/H] & \num{-0.78 \pm 0.17} & \num{-0.80 \pm 0.15} & \num{-0.71 \pm 0.16} & \num{-0.50\pm 0.16} \\
$V_\text{gas} - V_\ast$ [\si{\kilo\metre\per\second}] & \num{19.7 \pm 8.8} & \num{25.7 \pm 7.7} & \num{25.0 \pm 8.1} & \num{26.6 \pm 8.2} \\
\enddata
\tablecomments{The median and standard deviation of the bootstrapped values of the stellar population histograms in Figure~\ref{bootstrap_hist}. The last row reports the velocity difference between the gas and stellar velocities by calculating the median and standard deviation of $V_{\text{gas}} - V_\ast$.}
\end{deluxetable*}

The stellar velocity is approximately constant at \SI{230}{\kilo\metre\per\second} after an initial decrease of \SI{5}{\kilo\metre\per\second} from \ang{;;5} to \ang{;;10} apertures. The higher stellar velocity in the \ang{;;5} aperture is statistically consistent with the velocities measured in the larger apertures. The gas velocities remain constant at \SI{255}{\kilo\metre\per\second}. Since we measure these from \hii\ regions located in the disk, we take this velocity as the radial velocity of the disk.

Meanwhile, the stellar velocity dispersion increases as the aperture size increases. While \ppxf\ does automatically remove the instrumental dispersion, there are still contributions from the disk as well as the line-of-sight velocity difference between the \ac{bulge} and disk. Given the \ac{bulge} light fractions in Table~\ref{bootstrap_table}, the velocity dispersion measured in the \ang{;;5} aperture is likely most characteristic of the true velocity dispersion of the \ac{bulge}, although it is likely an upper limit (see also the discussion in \citealt{cappellari2017}). 

In the case of the gas velocity dispersion, there are a large number of bootstrapped solutions that give a dispersion of $\sim$\SI{1}{\kilo\metre\per\second}. This is likely the result of a ``failure mode'' in \ppxf\ where the gas velocity dispersion is unresolved in the spectra. Given the large FWHM of the V300 grating ($\sim$\SI{10}{\angstrom} or $\sim$\SI{460}{\kilo\metre\per\second} at \ha), larger than the internal velocities of \hii\ regions \citep[e.g.,][]{garciavazquez2023}, this is not surprising. Therefore, we show the histograms in Figure~\ref{bootstrap_hist} but do not report any median values. We checked whether these ``failed'' solutions had any effect on the gas velocities and they did not. 


Both the light- and mass-weighted ages and metallicities show variations with aperture size. As the aperture grows, both ages become younger, while the light-weighted metallicities remain constant at $\sim -0.5$ and the mass-weighted metallicities become more metal-rich although still subsolar. The constant light-weighted metallicities is understood to be dominated by the \ac{bulge} since that dominates the light profile at all aperture sizes (Table~\ref{bootstrap_table}). Meanwhile, the mass-weighted metallicities change with aperture size because at large apertures, the disk contributes more mass, leading to younger ages and more metal-rich populations. Interestingly, the secondary peak in stellar metallicity at $\text{[M/H]} \sim -0.4$ is remarkably similar to the gas-phase metallicity measured in \hii\ regions in this galaxy by \cite{moustakas2010}.


Finally, we can measure the velocity difference between the gas and stellar components. The median and standard deviations of $V_{\text{gas}} - V_\ast$ in each aperture is reported in Table~\ref{bootstrap_table}. Averaging over the four apertures, we find a velocity difference of $\simeq$\SI{24}{\kilo\metre\per\second} between the stellar absorption lines and the gas emission lines. We have also checked whether $V_{\text{gas}} - V_\ast$ changes if we measure the differences in the individual median values and propagate the uncertainties. This is identical within the uncertainties as expected for an approximately Gaussian distribution (Figure~\ref{bootstrap_hist}).


\begin{figure*}
\epsscale{0.9}
\includegraphics[keepaspectratio,width=\textwidth]{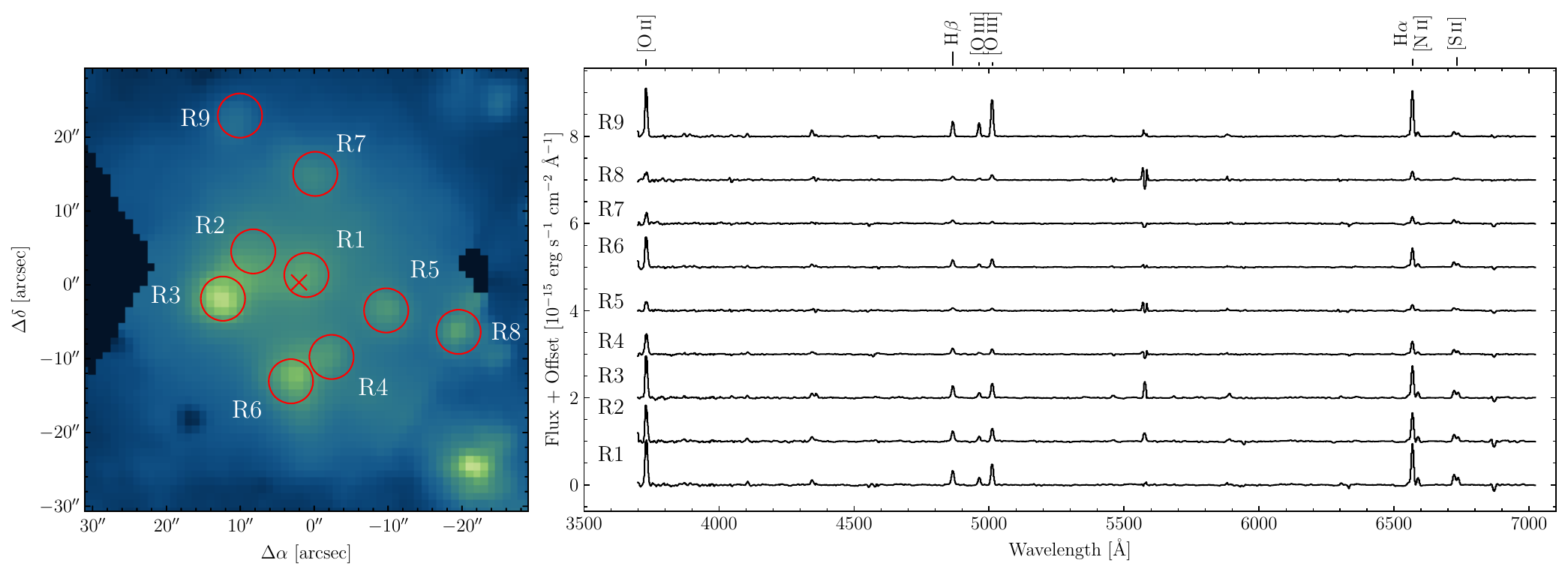}
\caption{Nine \hii\ regions in the region of the \ac{bulge}. The left panel shows a narrowband image selected from the data cube centered on the \ha\ emission line with a width of \SI{22}{\angstrom}. Circled in red and labeled are the nine selected \hii\ regions. The circles measure \ang{;;3} in radius. The red cross marks the center of NGC~5474. The right panel shows the spectra of the \hii\ regions. The strong lines used in the fitting process are labeled at the top of the panel.}
\label{hii_regions}
\end{figure*}

As a consistency check on the gas velocities measured by \ppxf, we also measured the individual strong lines' wavelengths in the disk's \hii\ regions. Figure~\ref{hii_regions} shows the locations and spectra in \ang{;;3} apertures of nine \hii\ regions in the \ac{bulge} region of NGC~5474. After subtracting the stellar continuum using \ppxf\ as described above, we fit a model consisting of nine Gaussians simultaneously to the strongest emission lines: \oii$\lambda$3727, \hb, \oiii$\lambda\lambda$4959,5007, \nii$\lambda\lambda$6548,6583, \ha, and \sii$\lambda\lambda$6717,6731. In addition to tying the central wavelengths of the lines together by their rest wavelengths, we also tie the widths of the oxygen, nitrogen, and sulfur doublets to be the same for each ionic species. 

The resulting average gas velocity is \SI{270.8}{\kilo\metre\per\second}. This methodology accurately measures the velocity of an individual \hii\ region to \SI{3.8}{\kilo\metre\per\second} similar to the uncertainties measured with \ppxf. However, there is mild scatter in the velocities of the \hii\ regions of \SI{7.3}{\kilo\metre\per\second} which contributes to the higher average gas velocity. This result agrees with the \hi\ disk velocity near the \ac{bulge} of $\sim$\SI{268}{\kilo\metre\per\second} \citep{rownd1994} and is in broad agreement with the gas velocities from \ppxf\ in that the gas velocities are larger than the stellar velocities. The mild disagreement with the gas velocities directly from \ppxf\ is likely caused by the large apertures including any diffuse or extended gas component in the average, blurring the contribution from any individual \hii\ region. Using the gas velocity measured from individual \hii\ regions, the average velocity difference is $\sim$\SI{39}{\kilo\metre\per\second}. As we shall see in the next section, this is consistent within the scatter seen in the spatially resolved velocity maps.

\subsection{2D Velocity Maps}

The second method for measuring the radial velocity uses our full \ac{ifs} coverage to produce 2D velocity maps. This technique complements and extends the aperture method, allowing us to investigate spatial variations in the disk and \acl{bulge}. Other 2D kinematic studies have found that typical spiral galaxies show no abrupt change in the stellar kinematics across a central bulge beyond what is expected in a rotating system (see many of the velocity maps in, e.g., \citealt{falconbarroso2006,garcialorenzo2015,falconbarroso2017,oh2022}). Additionally, there is usually less than a $\sim$\SI{2}{\kilo\metre\per\second} velocity difference between the stellar and gas velocities within the bulge \citep{martinsson2013}. Therefore, any significant variations in the stellar kinematics alone or between the gas and stellar kinematics in the disk and \ac{bulge} should provide key insights into the nature of the \ac{bulge}. 

Again, we use \ppxf, as described at the beginning of this section, to extract physical properties from the 2D \ac{ifs} data. In order to speed up computation time and improve the S/N in each pixel, we spatially bin our data into \qtyproduct[product-units=single]{3 x3}{\pixel} ($\ang{;;3} \times \ang{;;3}$) bins. We do not split up the stellar continuum and emission line fitting; instead, we opt to fit them simultaneously in one run of \ppxf. Note that we still mask the emission lines while fitting the stellar continuum. Finally, we do not bootstrap the model solutions. In its place, we apply regularization, which reduces the noise in the model fit and physical parameters (see Section~3.5 of \citealt{cappellari2017} for details).

As with the circular apertures, we extract the stellar and gas kinematics and the mass-weighted ages and metallicities in each binned pixel. We checked the reduced $\chi^2$ values of each model fit, and they were all approximately 1 to 2, indicating that the models are well-fit. Similarly, the S/N of each binned pixel was estimated. As expected, given our pointings and light distribution of NGC~5474, the S/N is highest in the \ac{bulge} with a median of $\sim$35 and decreases further into the disk. Thus, the properties located near or at the \ac{bulge} are likely well-determined and have lower uncertainties than those measured in the disk. We subsequently applied a S/N cut on the data and only kept data with a $\text{S/N} > 5$. This reduces the total coverage to the \ac{bulge} and some portions of the disk north and south of the \ac{bulge} as seen in the following discussion.


\begin{figure*}
\includegraphics[keepaspectratio,width=\textwidth]{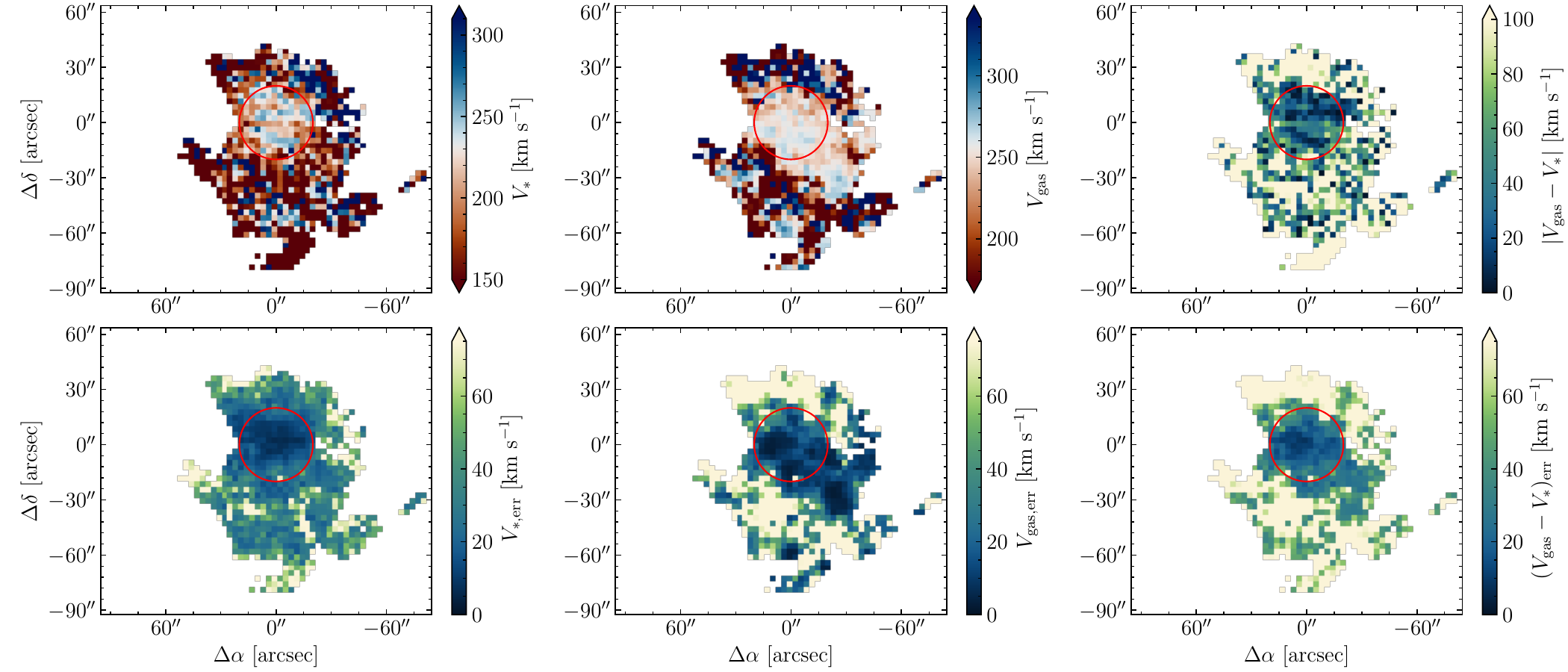}
\caption{From left to right: the stellar velocities, gas velocities, and velocity differences across NGC~5474. The top row of panels shows the velocities, while the bottom row shows their associated uncertainties. We masked areas of the galaxy where the $\text{S/N} < 5$. The \ang{;;20} red circle outlines the position of the \acl{bulge} in each map centered at the origin.  North is up and east is to the left.}
\label{vel-maps}
\end{figure*}

Figure~\ref{vel-maps} shows, from left to right, the stellar velocities, gas velocities, and velocity differences, along with the uncertainties for each quantity. The \ac{bulge} (circled in red) stands out from the disk with $V_\ast \simeq \SIrange{220}{240}{\kilo\metre\per\second}$, while clear clumps of \hii\ regions stand out compared to the disk with approximately the same velocities, $V_{\text{gas}} \simeq \SI{250}{\kilo\metre\per\second}$. These combine to make the \ac{bulge} stand out in the velocity difference map in the $|V_{\text{gas}} - V_\ast| \simeq \SIrange{20}{40}{\kilo\metre\per\second}$ range. This result is comparable to the velocity differences measured in the circular apertures in the previous section and strongly supports the idea that the \ac{bulge} is separate from the disk.

\begin{figure}
\includegraphics[keepaspectratio,width=\columnwidth]{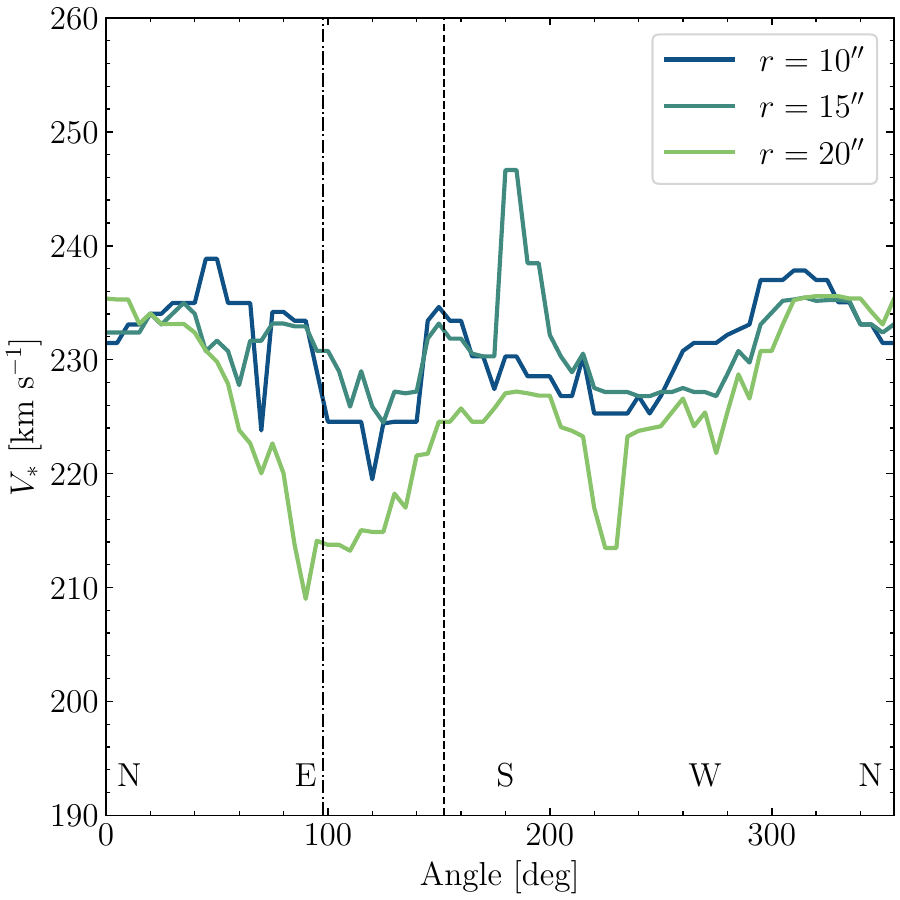}
\caption{The measured stellar velocities in a quadrant rotated in \ang{5} increments. Angles are measured east from north counterclockwise. Different colors measure the velocities within different aperture sizes indicated in the legend. Characteristic uncertainties for the velocities are \qtylist[list-units=single]{9.0;10.3;12.8}{\kilo\metre\per\second} for the \ang{;;10}, \ang{;;15}, and \ang{;;20} apertures, respectively. The vertical dashed line is the \hi\ kinematic position angle (\ang{158.5}; \citealt{rownd1994}) and the vertical dash-dot line is the photometric position angle (\ang{98}; \citealt{jarrett2003}).}
\label{vel-azi}
\end{figure} 

Given the higher S/N of the data in the \ac{bulge} region, we attempt to search for any signs of rotation in the \ac{bulge}, especially along the disk kinematic axis.\footnote{Due to the low S/N further out in the disk, we cannot make any robust measurement of disk rotation similar to the \hi\ data presented in \cite{rownd1994}.} The stellar kinematics map in Figure~\ref{vel-maps} does not appear to show any changing velocities across the \ac{bulge} region. Similar to the technique used in \cite{garner2022}, we measure the azimuthal changes in stellar velocities within \ang{;;10}, \ang{;;15}, and \ang{;;20} of the \acg{bulge} center in quadrants advanced incrementally by \ang{5}.\footnote{We do not use a \ang{;;5} aperture due to low number statistics with these binned pixels.} The results are shown in Figure~\ref{vel-azi}. We have also marked the position angles of the photometric major axis (dash-dot line; \citealt{jarrett2003}) and the \hi\ kinematic major axis (dashed; \citealt{rownd1994}). 

There does not seem to be any clear preference for rotation around either marked axis, nor rotation around any other axis. There is a hint of a signal in the \ang{;;20} aperture, but the disk contributes more to this aperture so this may be a measurement of disk rotation and not \ac{bulge} rotation. Unfortunately, a direct comparison with \hi\ disk rotation curves is not feasible since their resolution precludes measurements at the small radii presented here \citep{rownd1994,kornreich2000}. Extrapolating the rotation curve presented by \cite{kornreich2000} does suggest rotation signatures $\lesssim$\SI{10}{\kilo\metre\per\second}, likely due to the nearly face-on nature of NGC~5474. The lack of measured rotation does suggest that the \ac{bulge} is pressure-supported, but higher resolution data to estimate velocity dispersion more accurately is warranted to make that conclusion.


\begin{figure*}
\includegraphics[keepaspectratio,width=\textwidth]{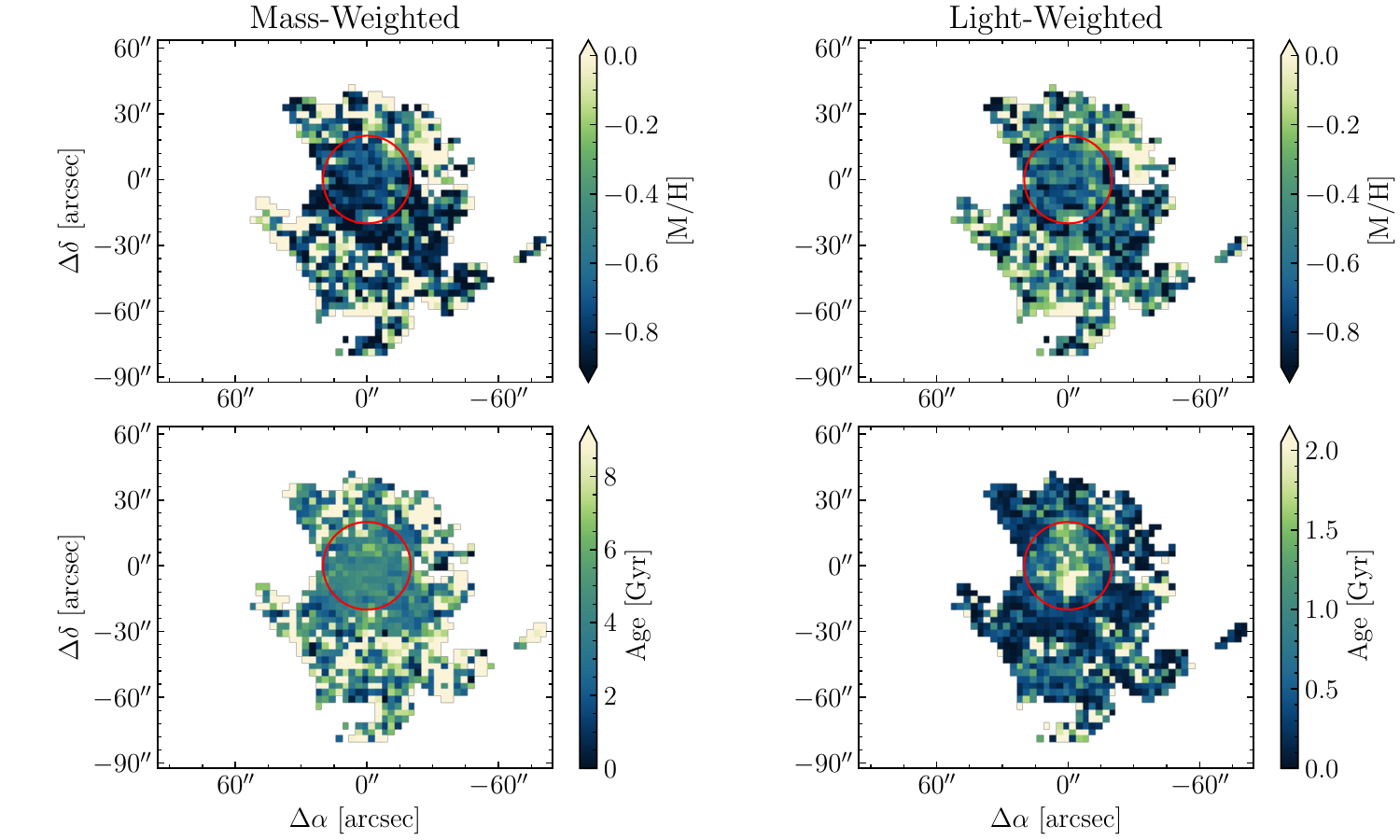}
\caption{The mass-weighted (left column) and light-weighted (right column) stellar metallicities (top) and ages (bottom) in the field-of-view. Note the different color bars for the stellar ages. Areas of the data with $\text{S/N} < 5$ were masked. The \ang{;;20} red circle outlines the position of \acl{bulge} in each map centered at the origin. North is up and east is to the left.}
\label{age_metal}
\end{figure*} 

Figure~\ref{age_metal} shows the mass- and light-weighted stellar metallicities and ages across the FOV. We again masked those binned pixels with $\text{S/N} < 5$. We estimate the median mass-weighted metallicity and age of the \ac{bulge} in a \ang{;;20} as \num{-0.72 \pm 0.23} and \SI{4.1 \pm 1.5}{\giga\year}; the median light-weighted metallicity and age of the \ac{bulge} are \num{-0.63 \pm 0.22} and \SI{0.8 \pm 0.7}{\giga\year}. For comparison, applying the stellar mass of the \ac{bulge} of \SI{5 \pm 0.3 e8}{\solarmass} \citep{bortolini2024} to the stellar mass-metallicity relation of \cite{kirby2013} predicts a stellar abundance of \num{-0.88 \pm 0.07}, entirely consistent with our estimated abundances. Thus, the \ac{bulge} is likely a low-mass object consistent with the scaling relations for dwarf galaxies. Unfortunately, the low S/N beyond the bulge precludes a robust measurement of the disk metallicity and age. Higher resolution spectroscopic data of the stellar component of NGC~5474, especially of the disk, is likely needed to make a detailed statement about the metallicities and ages of both components.

\section{Summary \& Implications}

Using data from the \ac{pings}, we have successfully and robustly measured a kinematic offset of $\simeq$\SI{24}{\kilo\metre\per\second} between the disk and \acl{bulge} of NGC~5474. This measurement, combined with a wealth of observations \citep[e.g.,][]{rownd1994,mihos2013,bellazzini2020,bortolini2024}, strongly suggests that the \ac{bulge} of NGC~5474 is not a bulge in the same plane as the disk, but instead a separate object seen along the line-of-sight. In the following, we attempt to put the puzzle pieces together, putting the velocity fields, photometry, and simulations in the same context. 

Our interpretation that the disk and the \ac{bulge} are separate objects draws on two key observations. The first is the striking velocity offset between the gas and stars in the region of the \ac{bulge}. In a survey of 30 isolated spiral galaxies, \cite{martinsson2013} compared the ionized gas and stellar velocity offsets. They found a median difference of $V_{\text{gas}} - V_\ast = \SI{-2.06 \pm 0.20}{\kilo\metre\per\second}$ such that the gas velocity is slightly blueshifted compared to the stars which \cite{martinsson2013} interpret as due to the near-side expansion of \hii\ regions. The velocity difference in NGC~5474 is an order of magnitude larger than that seen in typical spiral galaxies and in the opposite sense--the gas is redshifted with respect to the stars rather than blueshifted. Furthermore, NGC~5474 is not a strongly star-bursting galaxy ($\text{SFR} \simeq \SI{0.08}{\solarmass\per\year}$; \citealt{kennicutt2008}), lacks an AGN, and the \hi\ disk shows no velocity distortions near the \ac{bulge} \citep{rownd1994}. Thus, it is very unlikely that the velocity difference observed is caused by an outflow. 

The second observation is the apparent blueshift of the \ac{bulge} compared to the surrounding disk (Figure~\ref{vel-maps}). For isolated spiral galaxies, stellar velocity maps show smooth, regular rotation with no kinematic offset of the bulge (see many of the velocity maps in, e.g., \citealt{falconbarroso2006,ganda2006,martinsson2013,garcialorenzo2015,falconbarroso2017,guerou2017,oh2022}) unlike what is observed for NGC~5474. Could the velocity offset be a sign that the \ac{bulge} is currently merging with the disk of NGC~5474? Galaxies in the late stages of merging do show distortions to their stellar kinematics \citep[e.g.,][]{barreraballesteros2015,bloom2017,nevin2021,moralesvargas2023}, but these result in stronger \hi\ kinematic irregularities than is observed for NGC~5474 \citep{rownd1994,kornreich2000}. Meanwhile, galaxies in the earliest pre-merger stages show regular velocity fields with one galaxy being redshifted from the other, such as the optically superimposed galaxy pair VV488 (see Figure~B2 in \citealt{barreraballesteros2015}), which is qualitatively similar to NGC~5474. Thus, it is more likely that these two objects, the \ac{bulge} and disk of NGC~5474, are physically separate. 

While likely separate, the disk and \ac{bulge} are probably physically associated with each other since the velocity difference between the disk (gas) and \ac{bulge} (stellar) velocities is less than the circular speed of NGC~5474 ($\simeq$\SI{44}{\kilo\metre\per\second}, assuming $i = \ang{21}$; \citealt{rownd1994,kornreich2000}). Thus, the \ac{bulge} is likely an external satellite galaxy moving around the disk of NGC~5474 rather than a chance projection of distant objects \citep{pascale2021}. Given the lack of optical tidal features \citep{mihos2013,garner2021} and distorted velocity fields characteristic of galaxies in the process of merging (e.g., \citealt{barreraballesteros2015} and references therein), this is probably a weak interaction rather than an actively merging system. Models suggest that NGC~5474 had a previous interaction with M101 $\sim$\SI{300}{\mega\year} ago \citep{linden2022}, and clearly that close passage did not unbind the \ac{bulge}-disk pair. Thus this is likely a somewhat bound pair undergoing a weak interaction. 

These findings provide a compelling case that the \acl{bulge} of NGC~5474 is not an intrinsic bulge but rather a distinct satellite galaxy in the early stages of interaction with the disk. This interpretation sheds light on the peculiar dynamics and structure of NGC~5474 and highlights the complex evolutionary pathways that galaxies can undergo. Future work, including higher-resolution spectroscopy of the disk and bulge, as well as simulations and dynamical studies incorporating these two separate bodies into the M101-NGC~5474 interaction \citep[e.g.,][]{linden2022}, will be essential to confirm the nature of this interaction and explore its role in shaping the observed properties of NGC~5474 and M101. Ultimately, unraveling the nature of this interaction will deepen our understanding of satellite-disk dynamics and the broader processes that drive galaxy evolution in group environments.

\begin{acknowledgments}
The authors would like to thank the referee for comments that improved the clarity of the paper. R.G.\ would like to thank Grace Olivier and Briana Wirag for helpful comments and insightful discussions. Data here reported were acquired at Centro Astron\'{o}mico Hispano Alem\'{a}n (CAHA) at Calar Alto operated jointly by Instituto de Astrof\'{\i}sica de Andaluc\'{\i}a (CSIC) and Max Planck Institut f\"{u}r Astronomie (MPG). Centro Astron\'{o}mico Hispano en Andaluc\'{\i}a is now operated by Instituto de Astrof\'{\i}sica de Andaluc\'{\i}a and Junta de Andaluc\'{\i}a. This research has made use of the NASA/IPAC Extragalactic Database (NED), which is funded by the National Aeronautics and Space Administration and operated by the California Institute of Technology. This research has made use of the Astrophysics Data System, funded by NASA under Cooperative Agreement 80NSSC21M00561.

\end{acknowledgments}

\facility{CAO:3.5m (PMAS:PPak)}

\software{\texttt{Astropy} v5.3.4 \citep{astropy1,astropy2,astropy3}, \texttt{Matplotlib} v3.7.5 \citep{hunter2007}, \texttt{NumPy} v1.26.4 \citep{harris2020}, \texttt{SciPy} v1.11.4 \citep{virtanen2020}, \texttt{cmcrameri} \citep{crameri2018}, \textsc{r3d} \citep{sanchez2006}, \textsc{e3d} \citep{sanchez2004}, \textsc{pingsoft} \citep{rosalesortega2011}, \ppxf\ v9.1.1 \citep{cappellari2004,cappellari2017,cappellari2023}, \texttt{lineid\_plot}\footnote{\url{https://github.com/phn/lineid_plot}}}

\bibliography{Bulge_5474.bib}
\bibliographystyle{aasjournal.bst}

\end{document}